# Plastic inorganic $Sn_2BiS_2I_3$ semiconductor enabled deformable and flexible electronic tongue for heavy metal detection

Qiao Wang[a,b], Pengyue Zhao[a,b,*]

[a] Center of Ultra-Precision Optoelectronic Instrumentation Engineering, Harbin Institute of Technology, Harbin 150001, China

[b] Key Lab of Ultra-Precision Intelligent Instrumentation, Ministry of Industry Information Technology, Harbin 150080, China

**Abstract:**

Deformable and flexible electronics have garnered significant attention due to their distinctive properties; however, their current applications are primarily limited to the thermoelectric domain. Expanding the range of these electronics and their application scope represents a pivotal trend in their development. In this work, a plastic inorganic semiconductor material, $Sn_2BiS_2I_3$, with a band gap of 1.2 eV was synthesized and fabricated into a three-electrode flexible and portable electronic tongue capable of detecting heavy metal elements. The electronic tongue device exhibits exceptional linearity and demonstrates resistance against interference from impurity ions. The linear regression equation is expressed as $Y=0.24+19.06X$, yielding a linear coefficient of approximately 0.96, and the detectable limit stands at around 1.1 ppb, surpassing the 2.0 ppb limit of the ICP-AES instrument. Furthermore, mechanical testing reveals the favorable plasticity of $Sn_2BiS_2I_3$, as evidenced by the absence of cracks during nanoindentation. The indentation hardness of $Sn_2BiS_2I_3$ is approximately 1.67 GPa, slightly exceeding the hardness of Cu, which is 1.25 GPa. This study expands the repertoire of deformable and flexible electronics, offering a new and exceptional choice for biomimetic tongue sensor materials.

**Introduction:**

Heavy metal pollutants, such as Cd and Pb, confer substantial environmental and organismal hazards. These contaminants are distinguished by their elevated toxicity and refractoriness to remediation measures. Within the framework of food chain propagation, even marginal Cd concentrations can precipitate the progressive accumulation of heavy metal contaminants within organisms, culminating in inexorable harm over temporal scales[1,2]. Furthermore, the annual prevalence of Cd-induced toxicity on a global scale contributes to aberrant neurodevelopment in neonatal progeny. The recurrent emergence of water pollution incidents instigated by heavy metal pollutants underscores the imperative exigency for punctilious detection and amelioration of such contamination.

Developing an economical and uncomplicated detection technique plays a crucial role in achieving heavy metal pollutant detection. Commonly employed methods for detection encompass atomic absorption spectrometry (A.A.S.)[3], inductively coupled plasma atomic emission spectrometry (ICP-AES)[4], and inductively coupled plasma mass spectrometry (ICP-MS)[5,6], among others. Although these methods offer benefits such as high monitoring precision and detection efficiency, certain issues cannot be overlooked, including costly instruments, complex sample pre-treatment, high operational expenses, and the requirement for extensive expertise. In contrast, electrochemical methods, aided by simple and portable instruments[7-9], enable the development of numerous highly sensitive and selective molecular detectors. Moreover, advanced electrochemical enrichment technology allows for the rapid detection of heavy metals without the need for intricate pre-treatment procedures. Among the electrochemical methods, anodic stripping voltammetry (A.S.V.) exhibits a remarkable capability to concentrate on the working electrode surface and achieve a lower detection limit. As a result, it is widely acknowledged as the most effective approach for detecting trace amounts of heavy metals in environmental samples[10,11], including clinical and industrial wastewater samples[12]. Currently, the primary electrodes used in A.S.V. for heavy metal element detection are mercury film and suspended mercury[13]. However, due to the toxicity of mercury, there is an urgent need for an environmentally friendly electrode material to replace mercury electrodes[14,15].

To solve this problem, we have developed a plastic inorganic $Sn_2BiS_2I_3$ semiconductor electrode[16]. The electrode has the advantages of flexibility, green and non-toxic, etc. Meanwhile, it can sense all visible light as a narrow band gap semiconductor, which is conducive to further preparation of the photoelectrochemical (P.E.C.) detector[17-20]. In this work, the utilization of the plastic inorganic semiconductor $Sn_2BiS_2I_3$ was effectively demonstrated in the development of a portable electronic tongue capable of detecting the heavy metals Cd and Pb. It is worth noting that the current mainstream plastic inorganic semiconductors, such as ZnS and $Ag_2S$[21,22], are primarily employed in thermal applications[23,24]. Therefore, the successful application of $Sn_2BiS_2I_3$ represents a significant innovation in this field. The present research highlights a promising and captivating avenue for the utilization of plastic inorganic semiconductors.

The crystal structure of $Sn_2BiS_2I_3$, a quaternary sulfide iodide compound, is depicted in Figure 1a, which belongs to the tetragonal system with the Cmcm space group and exhibits a layered band structure. The lattice constants for the crystal are as follows: $a$ = 4.3214 (9) Å, $b$ = 14.258 (3) Å, $c$ = 16.488 (3) Å; $A$ = 4.2890 (6) Å, $B$ = 14.121(2) Å, $C$ = 16.414 (3) Å. The $Sn_2BiS_2I_3$ possesses strong anisotropy and can be described as a three-dimensional network comprising parallel infinite ribbons of [$M_4S_2I_4$] (M = Sn, Bi) extending along the crystallographic $c$-axis. The synthesis of $Sn_2BiS_2I_3$ single crystal was accomplished by enhancing the established chemical vapor deposition method[16], as illustrated in Figure 1b. Specifically, equal parts of $BiI_3$ (4N Aladdin 1000mg) and approximately twice the amount of S (4N Aladdin 110mg) and Sn (4N Aladdin 400mg) were homogeneously mixed in a corundum crucible measuring 50mm×20mm×20mm. A silicon substrate was positioned vertically near the inner wall of the crucible. The covered crucible was then placed in a tubular quartz furnace within a larger quartz furnace. Figure S1 shows the temperature profile during the growth of the $Sn_2BiS_2I_3$ single crystal. The synthesized $Sn_2BiS_2I_3$ single crystal, as observed in Figure 1c and Figure S2 using scanning electron microscopy (SEM), exhibit a banded quasi-two-dimensional layered structure. The optical image of the $Sn_2BiS_2I_3$ single crystal is presented in Figure S3. Energy Dispersive Spectroscopy (EDS) analysis (Figure 4d1) confirms the presence of characteristic peaks corresponding to the elements Sn, Bi, S, and I, with an atomic percentage ratio of approximately 2:1:2:3, as detailed in Table S1 and Figure 1c. The element mapping analysis in Figure 1d reveals a uniform distribution of the four elements throughout the crystal, indicating the absence of phase separation in $Sn_2BiS_2I_3$. Moreover, XRD spectrum has been employed as an effective technique to investigate crystal structure. In Figure 1e, the XRD spectrum displays peak positions at 41.6°, 43.4°, 44.7°, 50.7° and 57.0°, corresponding to the crystal face of

(153), (202), (221), (242) and (049), respectively, proving that we have synthesized $Sn_2BiS_2I_3$ crystals. The photoluminescence (PL) spectrum of the synthesized $Sn_2BiS_2I_3$ single crystal, shown in Figure S4, confirms a band gap of 1.2 eV.

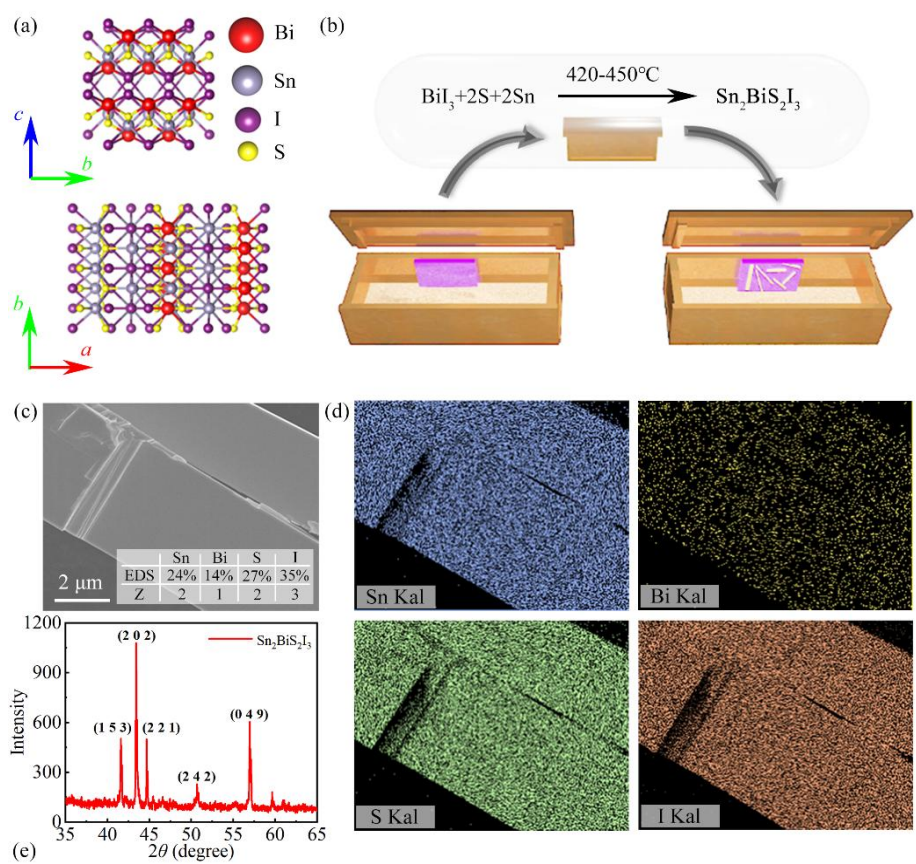

**Figure 1** Synthesis and characterization of $Sn_2BiS_2I_3$ single crystal. (a) Crystal structure, (b) Preparation process diagram, (c) SEM image of $Sn_2BiS_2I_3$, (d) Element mapping analysis, (e) XRD spectrum.

Figures 2a and 2b depict the differential charge density subsequent to the adsorption of Cd and Pb by $Sn_2BiS_2I_3$, with blue and yellow colors representing electron loss and gain, respectively. Based on electronic energy calculations of the geometric structure of $Sn_2BiS_2I_3$ after Cd and Pb adsorption, the adsorption energies for Cd and Pb on the surface of $Sn_2BiS_2I_3$ are determined as –0.313 eV and –1.797 eV, respectively. The disparity in adsorption energy can be quantitatively explained through differential charge density, and Bader charge analysis. For Cd adsorption, the differential charge analysis reveals the transfer of electrons from $Sn_2BiS_2I_3$ to Cd, with a transfer amount of 0.178 e determined through Bader charge calculations. Conversely, in the case of Pb adsorption, electrons transfer from Pb to $Sn_2BiS_2I_3$, with an electron transfer of 0.627 e, indicating a stronger interaction between Pb and the $Sn_2BiS_2I_3$ substrate, resulting in greater adsorption energy compared to Cd. As illustrated in Figures 2a and 2b, electron transfer between $Sn_2BiS_2I_3$ and Cd occurs from Cd to I, where I acts as the active site attracting Cd for bond formation. On the other hand, Sn and I contribute to the adsorption of Pb by $Sn_2BiS_2I_3$. From an atomic structure perspective, Pb possesses F-layer electrons. The weaker binding force between outer electrons and the Pb atomic nucleus, when compared to Cd, leads to greater electron transfer between Pb and Sn, resulting in higher adsorption energy. Figure 2c demonstrates the Electron Localization Function (ELF), with red indicating regions of high electron localization and blue representing the ionization region. The square-framed area in Figure 2c exhibits gradual electron localization following the adsorption of Pb and Cd, indicating an enhanced interaction between the adsorbed heavy metal elements and the $Sn_2BiS_2I_3$ substrate. The electron localization resulting from Pb adsorption is more pronounced than that from Cd, suggesting a certain degree of covalent effect, further elucidating the stronger adsorption of Pb by $Sn_2BiS_2I_3$. Figures 2d and 2e display the state density of $Sn_2BiS_2I_3$ before and after the adsorption of Cd and Pb. The geometric structures after the adsorption of Cd and Pb atoms can be observed in Figure S5. In the $Sn_2BiS_2I_3$ single crystal, the valence band predominantly consists of nonmetallic elements, namely S and I, while the conduction band is composed of metallic elements such as Bi. The partial wave density of states in Figure S6 reveals P orbital contributions from Bi, S, and I. Furthermore, a clear overlap between S/I and Sn/Bi orbitals indicates their hybridization. Adsorption of Cd on the surface of $Sn_2BiS_2I_3$ introduces impurity states within the band gap, serving as electron-hole recombination centers, which is unfavorable for exciton generation. Conversely, the incorporation of Pb induces a downward shift in the band edge, enhancing the oxidation ability of the valence band edge.

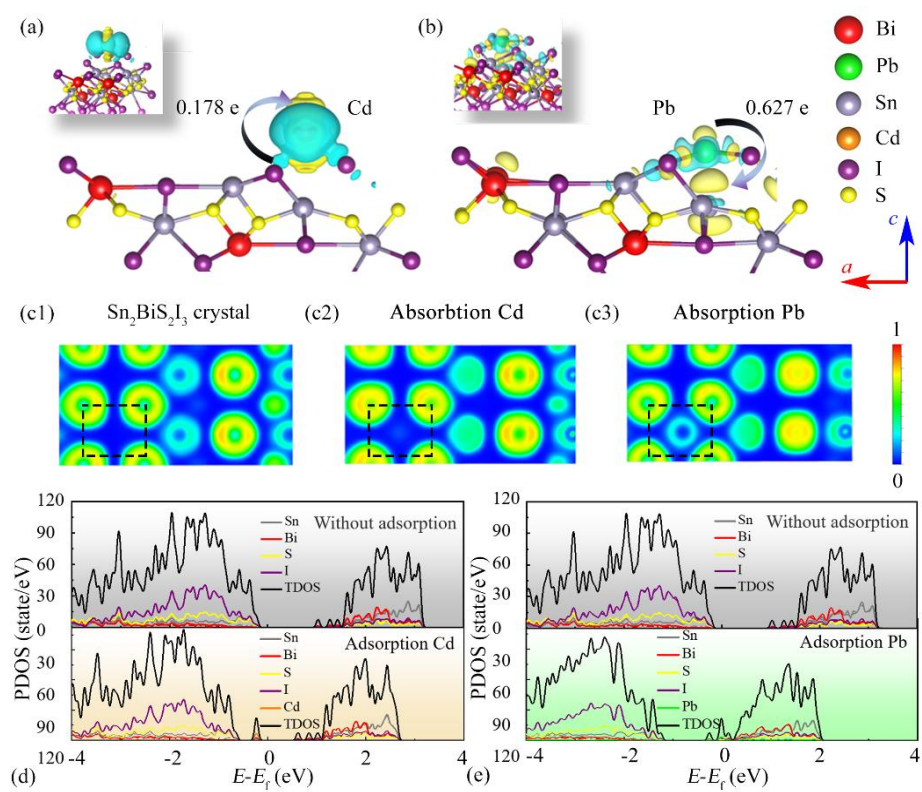

**Figure 2** Density functional calculation of heavy metals Cd and Pb adsorption on $Sn_2BiS_2I_3$ surface. (a) and (b) Differential charge density for adsorption Cd and Pb, respectively. (c) Electron localization function before and after adsorption. (d) The density of state after adsorption Cd. (e) The density of state after adsorption Pb.

The tongue serves as a vital organ enabling animals to perceive the surrounding environment, specifically engaging in the detection of certain distinctive molecules that elicit taste sensations[30]. Nevertheless, this biological detection mechanism exhibits evident limitations. The taste perception by the tongue is qualitative in nature, lacking the ability to precisely discern the quantity of sugar or salt present in a given cup of water. Furthermore, the scope of perception for the tongue is constrained, limited solely to molecules capable of producing taste, while certain toxic molecules prove challenging to detect through this means. As bionic robots continue to advance, there is an increasing demand for highly sophisticated electronic tongue sensors[31,32], which requires the ability of electronic tongues to surpass biological tongues.

In this work, the process of fabricating the plastic $Sn_2BiS_2I_3$-based electronic tongue is illustrated in Figure 3a. The electronic tongue device comprises a planar and flexible three-electrode system. Initially, the $Sn_2BiS_2I_3$ micron rod, grown on a silicon substrate, is delicately affixed using Scotch tape and subsequently transferred onto a PDMS substrate. Subsequently, the PDMS transfers the $Sn_2BiS_2I_3$ micron rod onto a commercially acquired flexible three-electrode system that lacks a working electrode. The three-electrode system incorporates an Ag/AgCl electrode as the reference electrode, an electron beam evaporation platinum electrode as the counter electrode, and $Sn_2BiS_2I_3$ as the working electrode for heavy metal detection. By utilizing an adhesive, the $Sn_2BiS_2I_3$ micron rod is securely affixed to the flexible three-electrode substrate, and the connection between the $Sn_2BiS_2I_3$ working electrode and silver glue and copper wire is established. The resultant device can be inserted into a commercial three-electrode socket, as depicted in Figure 3b.

By employing a pipette gun, a buffer solution comprising heavy metal constituents at a pH of 4.4 was introduced to the three electrodes of the $Sn_2BiS_2I_3$ electronic tongue. The detection of Cd ions, a representative heavy metal, was conducted through anodic stripping voltammetry[33], which encompasses two sequential stages: enrichment and dissolution, as illustrated in Figure 3c. By employing reduction potential adsorption and enrichment, followed by square-wave voltammetry (SWV) forward scanning dissolution, the oxidation peak of Cd can be acquired. Refer to the support letter for specific parameters. It is obvious from the EDS curve comparison in Figure 3d that cadmium can be enriched on the surface of $Sn_2BiS_2I_3$.

Figure 3e shows the voltammetric response curve of Cd (II) ions within the concentration range of 0.001 uM to 0.2 uM using a three-electrode electronic tongue pair. It illustrates the determination of the dissolution potential and peak strength of the $Sn_2BiS_2I_3$ electrode towards Cd (II) ions. Notably, $Sn_2BiS_2I_3$ exhibits a favorable voltammetric response towards Cd ions, with the electric signal increasing proportionally to the concentration of divalent cadmium ions. Furthermore, as depicted in Figure 3e illustration, the standard curve of the $Sn_2BiS_2I_3$ electronic tongue for detecting Cd (II) demonstrates a robust linear relationship between the magnitude of the peak current and the Cd (II) concentration. The linear regression equation is $Y=0.24+19.06X$, with a linear coefficient of 0.96. Experimental tests for the detection of cadmium ions at various concentrations indicate that the minimum detectable limit of the $Sn_2BiS_2I_3$ electronic tongue is approximately 0.01 uM, equivalent to 1.1 ppb based on the relative atomic weight of cadmium. Notably, this surpasses the 2 ppb detection limit of Cd (II) achieved by the ICP-AES instrument.

Figure 3f illustrates the heteroion resistance capability of the $Sn_2BiS_2I_3$ electronic tongue. The solution under examination is supplemented with common metal ions, namely Na (I), Mg (II), Ca (II), as well as the heavy metal element Pb in water. The square wave voltammetry (SWV) plot of the mixed ion solution demonstrates the absence of interference dissolution potential for Pb and Cd ions. This observation indicates that the $Sn_2BiS_2I_3$ electronic tongue possesses the ability to detect Cd, Pb, and other heavy metals in practical scenarios. The $Sn_2BiS_2I_3$ electronic tongue offers notable cost advantages and exceptional portability compared to the costly ICP (inductively coupled plasma) and AAS (atomic absorption spectrometry) instruments. To evaluate the precision of the $Sn_2BiS_2I_3$ electronic tongue, a comparison was made between its measurements and those obtained using the

ICP-OES (inductively coupled plasma optical emission spectrometer) in the Majiagou River. The test outcomes indicate a mere 7% disparity between the two measured values (Figure S7).

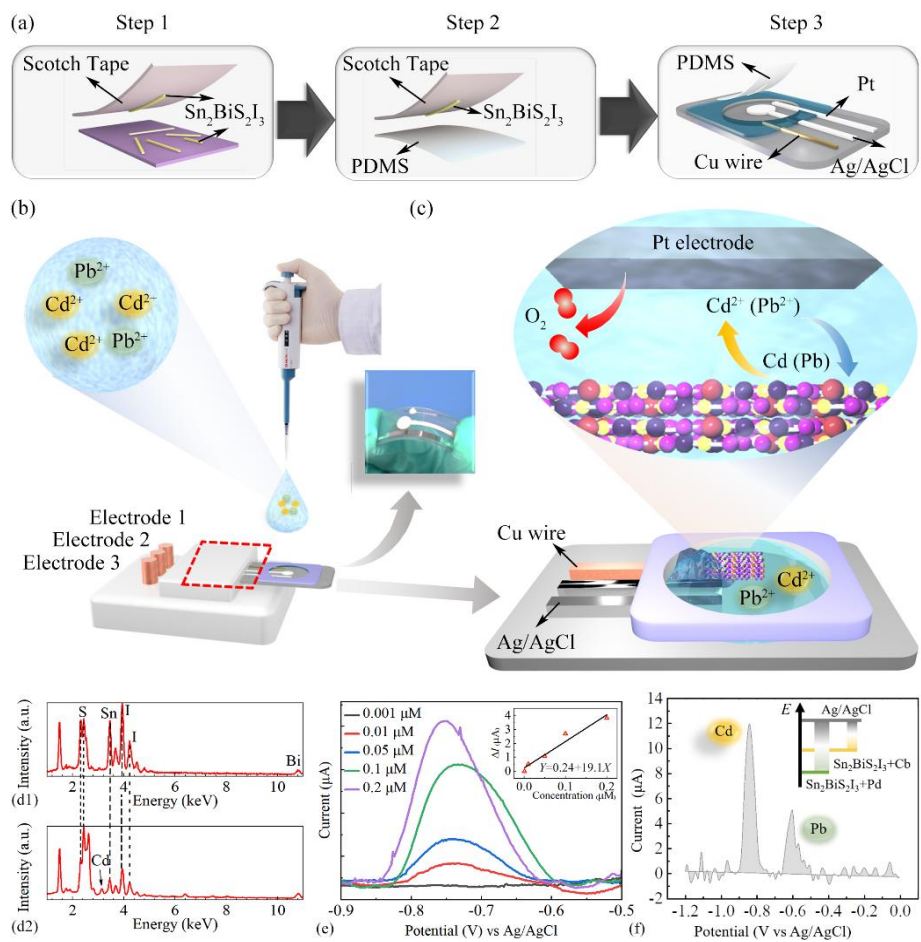

**Figure 3** Preparation process of electronic tongue and its performance for heavy metal detection. (a) The preparation process of $Sn_2BiS_2I_3$ electronic tongue, (b) Photo of electronic tongue and schematic diagram of the detection process, (c) Device details and working principle. (d) The EDS curve before (d1) and after (d2) detection of heavy metal Cd (II). (e) The SWV square wave voltammetric curve at different concentrations and the illustration is the standard curve of $Sn_2BiS_2I_3$ electronic tongue for detecting Cd (II) (f) Square wave voltammetric curve containing interfering ions.

The mechanical properties of $Sn_2BiS_2I_3$ single crystal were tested by G200 nanoindentation machine, and the test diagram was shown in Figure 4a and 4d, encompassing three stages: loading, stabilization, and unloading (Figure S8). Figure 4b and Figure S9a reveal the absence of any surface cracks on the $Sn_2BiS_2I_3$ single crystal, indicative of its remarkable plastic deformation capability. Figure S9b helps to demonstrate its excellent plasticity, with no cracks in the case of bending. Figure 4c displays the Focused Ion Beam (FIB)-prepared indentation section sample, demonstrating an indentation depth of approximately 300 nm without any underlying cracks. Furthermore, Figure 4e, presenting transmission electron microscopy (TEM) results, exhibit bending deformation behavior in the nanoindentation of the $Sn_2BiS_2I_3$ single crystal with its two-dimensional layered structure, akin to the characteristics observed in InSe material as reported in the previous literature[28]. This behavior is likely responsible for the exceptional plasticity observed in $Sn_2BiS_2I_3$ single crystal. To assess the plasticity of $Sn_2BiS_2I_3$ single crystal materials under nanoindentation, we performed nanoindentation experiments on GaN, $Bi_2Se_3$ and Cu, $MoS_2$, representative of brittle and plastic materials, respectively, using identical parameters. Figure 4f demonstrates the absence of pop-in phenomena during the indentation process of $Sn_2BiS_2I_3$ and Cu, indicating the absence of brittle crack formation in $Sn_2BiS_2I_3$. Conversely, the load-depth curve of GaN and $Bi_2Se_3$ exhibits a conspicuous pop-in phenomenon, similar to ZnS[28], indicating the brittleness of GaN and $Bi_2Se_3$. It is worth noting that the nanoindentation performance of $Sn_2BiS_2I_3$ is much better than that of the classical plastic van der Waals layered material $MoS_2$, which has a serious pop-in phenomenon.

To ensure reliable results, five nanoindentation experiments were conducted on Cu, $MoS_2$, $Bi_2Se_3$, GaN, and $Sn_2BiS_2I_3$, respectively. The findings reveal that the hardness and elastic modulus of GaN are substantially higher than those of other materials, and $Sn_2BiS_2I_3$ is between GaN, $Bi_2S_3$ and $MoS_2$, Cu, as shown in Figure 4g and 4h. The modulus of the $Sn_2BiS_2I_3$ single crystal is approximately 1.50 GPa, which is smaller than that of $MoS_2$ 1.24 GPa. Moreover, the hardness of the $Sn_2BiS_2I_3$ single crystal exceeds that of $MoS_2$, measuring approximately 121.97 GPa. Thus, the $Sn_2BiS_2I_3$ single crystal exhibits excellent plasticity.

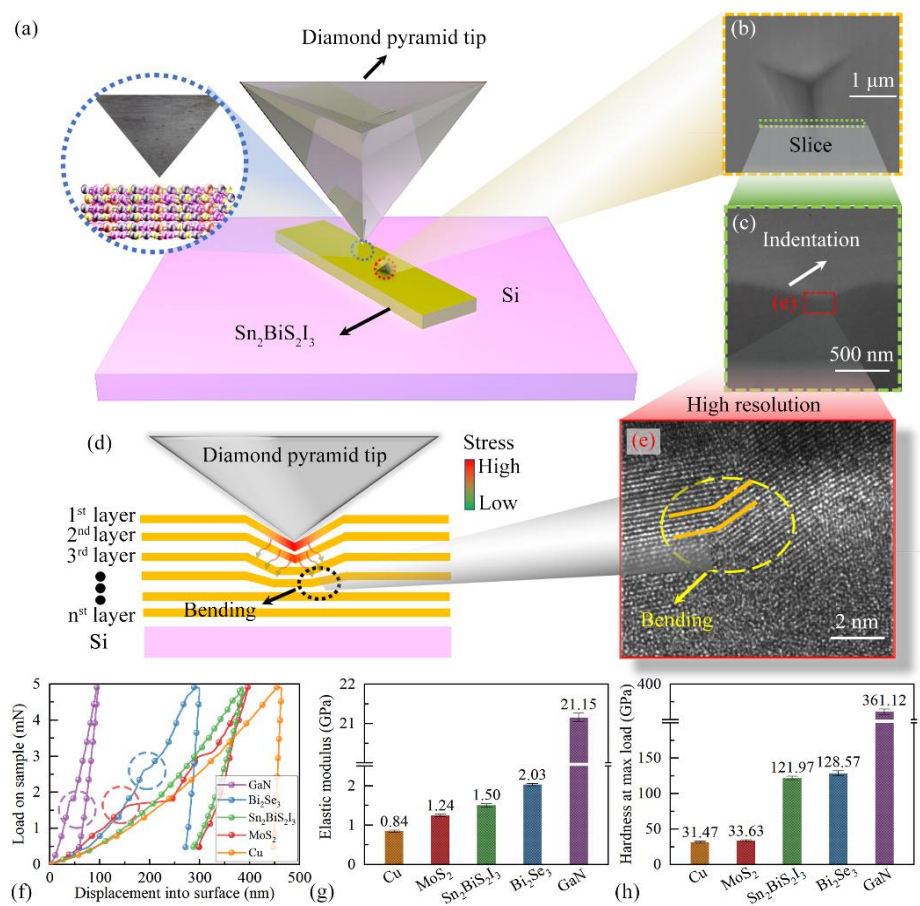

**Figure 4** Mechanical properties of $Sn_2BiS_2I_3$ single crystal. (a) Schematic diagram of nanoindentation, (b) SEM image of indentation surface, (c) Indentation section obtained by FIB, (d) The deformation mechanism for $Sn_2BiS_2I_3$ single crystal, (e) TEM high-resolution image for $Sn_2BiS_2I_3$ in plastic deformation area, (f) Load-displacement curve for $Sn_2BiS_2I_3$, Cu and GaN, (g) Elastic modulus for $Sn_2BiS_2I_3$, Cu and GaN, (h) Hardness for $Sn_2BiS_2I_3$, Cu and GaN.

This work presents the characterization of the plastic inorganic $Sn_2BiS_2I_3$ semiconductor material. Nanoindentation test results reveal that $Sn_2BiS_2I_3$ possesses a modulus akin to $MoS_2$ yet significantly lower than that of GaN. Moreover, a sensory device utilizing $Sn_2BiS_2I_3$, referred to as an electronic tongue sensor, effectively detects heavy metal elements such as Cd and Pb. The intricate structure and composition of Sn$_2$BiS$_2$I$_3$, in comparison to simpler binary plastic semiconductors, impart promising potential for enhancing its mechanical and electrical properties. Furthermore, Sn$_2$BiS$_2$I$_3$ exhibits a unique one-dimensional layered structure, distinguishing it from two-dimensional layered plastic semiconductors. Although there are ongoing challenges in fabricating bulk Sn$_2$BiS$_2$I$_3$ materials, the successful integration of deformable and flexible inorganic materials into the sensor domain represents groundbreaking work, establishing a robust foundation for the diverse applications of such materials.

**Supporting Information**
Supporting Information is available from the Wiley Online Library or from the author.

**Acknowledgments**
This research was funded by the National Key Research and Development Program of China (2022YFF060503 and 2023YFF0612501), the Young Elite Scientists Sponsorship Program by CAST (2022QNRC001), the National Natural Science Foundation of China (52305569, 52205560, 52175499 and 52105547), the Open Project Program of State Key Laboratory of applied optics (SKLAO2021001A05) and the State Key Laboratory of applied optics.

**Conflict of Interest**
The authors declare no conflict of interest.